\documentclass[aps,prl,showpacs,superscriptaddress,amsmath,twocolumn]{revtex4-1}

\newcommand\supplementPRL{%
\setcounter{equation}{0}%
\setcounter{figure}{0}%
\renewcommand \theequation{S\arabic{equation}}%
\renewcommand \thefigure{S\arabic{figure}}%
}

\usepackage{amsmath}
\usepackage{amssymb}
\usepackage{graphicx}

\usepackage{epsfig}

\usepackage{times}

\begin{document}

\title{Effect of the band structure topology on the minimal conductivity for bilayer graphene with symmetry breaking }

\author{Gyula D\'avid
}
\affiliation{Department of Atomic Physics,
E{\"o}tv{\"o}s University,
H-1117 Budapest, P\'azm\'any P{\'e}ter s{\'e}t\'any 1/A, Hungary}

\author{ P\'eter Rakyta
}
\affiliation{Department of Physics of Complex Systems,
E{\"o}tv{\"o}s University
\\ H-1117 Budapest, P\'azm\'any P{\'e}ter s{\'e}t\'any 1/A, Hungary}

\author{ L\'aszl\'o Oroszl\'any
}
\affiliation{Department of Physics of Complex Systems,
E{\"o}tv{\"o}s University
\\ H-1117 Budapest, P\'azm\'any P{\'e}ter s{\'e}t\'any 1/A, Hungary}

\author{ J\'ozsef Cserti
}
\affiliation{Department of Physics of Complex Systems,
E{\"o}tv{\"o}s University
\\ H-1117 Budapest, P\'azm\'any P{\'e}ter s{\'e}t\'any 1/A, Hungary}

\begin{abstract}

Using the Kubo formula we develop a general and simple expression for the minimal conductivity in systems described
by a two by two Hamiltonian.
As an application we derive an analytical expression for the minimal conductivity tensor of bilayer graphene
as a function of a complex parameter $w$ related to recently proposed symmetry breaking mechanisms resulting  
from electron-electron interaction or strain applied to the sample.
The number of Dirac points changes with varying parameter w, this
directly affect the minimal conductivity.
Our analytic expression is confirmed using an independent calculation based on Landauer approach and we find remarkably good agreement between the two methods.
We demonstrate that the minimal conductivity is very sensitive to the change of the parameter $w$ and the orientation of the electrodes with respect to the sample.
Our results show that the minimal conductivity is closely related to the topology of the low energy band structure.

\end{abstract}






\maketitle

\emph{Introduction.}---After the first quantum Hall measurement on graphene~\cite{Novoselov_graphene-1,Zhang_graphene:ref}
the physics of graphene has become one of the leading research field in physics.
The bilayer graphene has been studied first experimentally~\cite{Novoselov_Hall:ref} by Novoselov \textit{et al.} and
theoretically~\cite{mccann:086805} by McCann and  Fal'ko.
Recent observations~\cite{ISI:000273086700016,ISI:000271655100037} indicate that in bilayer graphene spontaneous symmetry breaking
may arise from electron-electron Coulomb interactions.
Occurance of such broken-symmetry states generates more attention to the topological changes in Fermi surface in high quality
suspended bilayer graphene devices.
Lemonik \textit{et al.} studied the spontaneous symmetry breaking and Lifshitz transition in bilayer graphene~\cite{PhysRevB.82.201408}.
Spontaneous inversion symmetry breaking in graphene bilayers has also been investigated by Zhang \textit{et al.}~\cite{PhysRevB.81.041402}.
Vafek and Yang applied renormalization group approach to study the many-body instability of Coulomb interacting bilayer graphene~\cite{PhysRevB.81.041401}.
Nandkishore and Levitov   \cite{Levitov_flavour_PhysRevB.82.115124:cikk}, and Gorbar \textit{ et al.}~\cite{Gusynin_1_2:cikk}
studied competition between different ordered states in bilayer graphene.
Spontaneous symmetry breaking in two-dimensional electronic systems with a quadratic band crossing was studied
by Sun \textit{et al.}~\cite{PhysRevLett.103.046811}.
The quantum theory of a nematic Fermi fluid has been proposed by Oganesyan \textit{et al.}~\cite{PhysRevB.64.195109}.

The low energy Fermi surface of bilayer graphene is dominated by the trigonal warping, first shown by McCann and Fal'ko~\cite{mccann:086805}.
At very low energies, namely below the Lifshitz energy the trigonal warping results in a breaking of the constant energy lines into four pockets.
The Lifshitz energy is typically of order of $1$ meV~\cite{ISI:000249758000011}.
Besides many other effects the trigonal warping has a substantial influence on the minimal conductivity $\sigma^{\text {min}}$.

Without trigonal warping in Ref.~\onlinecite{Cserti_PhysRevB.75.033405} it was found that $\sigma^{\text {min}} = 8\, \sigma_0$, where
$\sigma_0 = e^2/(\pi h)$.
This later was confirmed by Snyman and Beenakker~\cite{PhysRevB.75.045322} using the Landauer approach.
Actually, the calculated minimal conductivity can take non-universal values depending on the order of the dc limit and the integration over energies
as shown by Ziegler~\cite{PhysRevB.75.233407}.
The importance of the order of the frequency and the temperature limit was pointed out by Ryu \textit{et al.}~\cite{Ryu-Ludwig:cikk_PhysRevB.75.205344}.
Trushin\textit{ et al.} showed that electron-hole puddle formation is not a necessary condition for finite conductivity in bilayer
graphene at zero average carrier density~\cite{PhysRevB.82.155308}.
Culcer and Winkler studied the role of the external gates and transport in biased bilayer graphene using the density operator formalism
and quantum Liouville equation ~\cite{PhysRevB.79.165422}.
From self-consistent Born approximation Koshino and Ando found that in the strong-disorder regime $\sigma^{\text {min}} = 8\, \sigma_0$,
while in the weak-disorder regime $\sigma^{\text {min}} = 24\, \sigma_0$.
In Ref.~\cite{cserti:066802} the role of the trigonal warping was studied and it was shown that the contributions of the four pockets
to the minimal conductivity gives $\sigma^{\text {min}} = 24\, \sigma_0$.
Moghaddam and Zareyan showed that the minimal conductivity of graphene bilayers is anisotropic with respect to
the orientation of the connected electrodes when the trigonal warping is taken into account~\cite{PhysRevB.79.073401}.

The spontaneous symmetry breaking in bilayer graphene causes changes in the low energy band structure,
namely the position of the pockets and even the number of pockets can alter.
The symmetry breaking induced by the electron-electron interaction in bilayer graphene can adequately be described
by the Hamiltonian suggested by Lemonik \textit{et al.}~\cite{PhysRevB.82.201408}.
Very recently, the same form of the Hamiltonian has been derived by Mucha-Kruczy\'nski,  Aleiner and Fal'ko
for electrons in strained bilayer graphene~\cite{Falko_strain_2layer:cikk}.
They studied the band structure topology and Landau level spectrum for strained bilayer graphene.
The Hamiltonian depends on a complex parameter $w$ and its change causes transition in the electronic band structure.
However, the effect of the symmetry breaking on the minimal conductivity is an open question.

In this work, we calculate the minimal conductivity as a function of the parameter $w$.
We find that the change of this parameter can dramatically affect on the minimal conductivity.
Starting from the Kubo formula we develop a general and simple method to find the minimal conductivity for a wide class of Hamiltonians.
Using this general approach we derive an analytical expression for the conductivity tensor
in bilayer graphene with symmetry breaking by complex parameter $w$.
As a self check we performed numerical calculations based on the Landauer formula and the agreement is very good.
Our analysis of the minimal conductivity presented in this work was inspired by recent experiment and insightful discussions
with Novoselov~\cite{Novoselov_private_comm}.

\emph{General approach.}---To calculate the conductivity of various two-dimensional electronic systems with electron-hole symmetry,
we consider a general model Hamiltonian:
\begin{eqnarray}
H(\mathbf{p}) &=&  \left( \begin{array}{cc}
0 & h({\bf p})\\
{h^*({{\bf p}})} & 0
\end{array}  \right) ,
\label{gen-Ham:eq}
\end{eqnarray}
where $h({\bf p}) = h_1({\bf p}) + i h_2({\bf p})$ and $*$ denotes the complex conjugation.
Specifically, the Hamiltonian for bilayer graphene including symmetry breaking has this form with
\begin{equation}
 h({\bf p}) = - \varepsilon_{\text{L}}  \left(\frac{p_-^2}{2} -  p_+ - \frac{w}{2} \right),
\label{hp_2layer:eq}
\end{equation}
valid for the valley  ${\bf K}$.
Here $\varepsilon_{\text{L}}=m v^2_3/2$ is the Lifshitz energy (where $m$ and $v_3$ are given in Ref.~\cite{mccann:086805}),
the dimensionless momentums are
$p_\pm = p_x \pm i p_y $ (in units of $m v_3$).
Finally, $w$ is generally a complex parameter (independent of $\mathbf{k}$) and can be originated
from the electron-electron interaction~\cite{PhysRevB.82.201408} and/or from the applied strain~\cite{Falko_strain_2layer:cikk}
and/or from the slide of the layer~\cite{Falko_strain_2layer:cikk,Son_slide_top_2layer:cikk}.


To find the conductivity for clean and bulk systems we start from the general Kubo formula presented in Ref.~\onlinecite{cserti:066802}
(derived from the form given by Ryu \textit{et al.}~\cite{Ryu-Ludwig:cikk_PhysRevB.75.205344}).
The minimal dc conductivity (at zero frequency $\omega$, at zero temperature and at zero Fermi energy) is given by
\begin{subequations}%
\label{sigma_Kubo:eq}
\begin{eqnarray}
\sigma^{\text{min}}_{lm} &=& n_d \,  \frac{2e^2}{h}\, \lim_{\eta\to 0}  I_{lm}(\eta), \label{sigma_min_lim:eq}\,\,\, \text{where}
\label{sigma-eta-2:eq}\\
I_{lm}(\eta) &=&  \eta^2  \int \frac{d^2 {\bf k}}{{\left(2\pi\right)}^2}\,
\text{Tr} \,  T_{lm}({\bf k},\eta ),  \label{Imunu:eq} \\
 T_{lm}({\bf k},\eta ) &=& {\left[\eta^2 + H^2({\bf k}) \right]}^{-1}
\frac{\partial H({\bf k})}{\partial k_l} \nonumber \\
&\times&
{\left[\eta^2 + H^2({\bf k}) \right]}^{-1}
\frac{\partial H({\bf k})}{\partial k_m},
\label{int-sigma:eq}
\end{eqnarray}%
\label{int-sigma-egybe:eq}%
\end{subequations}%
and ${\bf p}=\hbar {\bf k}$, $l,m=x,y$, and $n_d $ is the degeneracy (for bilayer graphene $n_d = n_s n_v$,
where $n_s =2$ is the spin degeneracy and $n_v =2$ is the valley degeneracy corresponding to
the valley ${\bf K}$ and ${\bf K}^\prime$).
The parameter $\eta$ can be interpreted physically as a finite inverse lifetime induced by impurities.

The eigenvalues of the Hamiltonian (\ref{gen-Ham:eq}) are $E({\bf k}) = \pm \sqrt{h^*({\bf k}) h({\bf k})}$.
Since $H^2 = E^2 I_2$ (here $I_2$ is the 2 by 2 unit matrix), the operator inverse in Eq.~(\ref{int-sigma:eq}) can be written as
${\left[\eta^2+H^2({\bf k})\right]}^{-1}  = {\left[\eta^2+E^2({\bf k})\right]}^{-1} \, I_2$.
Hence, the main contribution of the integrand  in (\ref{Imunu:eq}) comes from the vicinity of the zeros of the energy eigenvalues.
The real solutions of $E({\bf k} ) =0 $ are denoted by ${\bf k}^{(s)}$, where  $s = 1, \dots n_\text{D}$ and $n_\text{D}$ is the number of zeros
(for bilayer graphene $n_\text{D} =2,3,4$ depending on the parameter $w$).
Expanding the matrix $T_{lm}$ in (\ref{int-sigma:eq}) around one of the zeros ${\bf k}^{(s)}$, ie, for fixed $s$
the denominator becomes a polynomial of ${\bf q}$, where ${\bf q} =  {\bf k}-{\bf k}^{(s)}$.
If the energy dispersion $E({\bf k} )$ at the point ${\bf k}^{(s)}$ forms a Dirac cone
then the leading term of the expansion is quadratic in  ${\bf q}$ (otherwise one needs to expand $E^2(\bf k)$ beyond the quadratic terms in ${\bf q}$), 
and can be written as $E^2({\bf q}) = \sum_{i,j} M_{ij} \, q_i \, q_j$,
where the matrix $M$ is positive semi-definite.
For the Hamiltonian (\ref{gen-Ham:eq}) one finds
\begin{equation}
 M_{ij} =
\frac{\partial h_1}{\partial k_i}\, \frac{\partial h_1}{\partial k_j} + \frac{\partial h_2}{\partial k_i}\, \frac{\partial h_2}{\partial k_j},
\label{gen-M-matrix:eq}
\end{equation}
where the derivations are evaluated at ${\bf k} = {\bf k}^{(s)}$.
Moreover, we find that
$\text{Tr} \left[\frac{\partial H({\bf k})}{\partial k_l} \, \frac{\partial H({\bf k})}{\partial k_m}\right] = 2 M_{lm}$ and
$T_{lm}({\bf q},\eta) = 2 M_{lm}/{\left(\eta^2 + \sum_{i,j} M_{ij} \, q_i \, q_j \right)}^2$.

Now substituting this expression into (\ref{Imunu:eq}) and re-scaling the wave number
${\bf q}$ as ${\bf q} = \eta \, {\bf q}^\prime $ the integral $I_{lm}$ 
becomes\textit{ independent} of $\eta$, and reads
\begin{eqnarray}
 I_{lm} &=& 2 M_{lm}\, \int \frac{d^2 {\bf q^\prime}}{{\left(2\pi\right)}^2}\,
\frac{1}{{\left(1 + \sum_{i,j} M_{ij} \, q^\prime_i \, q^\prime_j \right)}^2}.
\label{I_mu_nu-2:eq}
\end{eqnarray}
Note that for the limit  $\eta \to 0 $ the main contribution in the integral  $I_{lm}$  comes from the vicinity of each Dirac points; 
therefore, the integral over ${\bf q}^\prime$ can be extended to infinity.

If the determinant of the matrix $M$ at ${\bf k}^{(s)}$ is zero then the integral is divergent.
For finite  determinant of $M$  the integral in (\ref{I_mu_nu-2:eq}), ie, the contribution to the integral (\ref{Imunu:eq})
over the vicinity of ${\bf k}^{(s)}$ can be performed analytically and it becomes $I^{(s)}_{lm} =  \frac{1}{2\pi}\, M^{(s)}_{lm}/\sqrt{\det M^{(s)}}$.
Here $M^{(s)}_{lm}$ denotes the matrix defined in Eq.~(\ref{gen-M-matrix:eq}) evaluated at ${\bf k} = {\bf k}^{(s)}$.

Then using Eq.~(\ref{sigma_min_lim:eq}) the minimal conductivity
is the sum of the contributions from each Dirac cones at ${\bf k}^{(s)}$:
\begin{equation}
 \sigma^{\text{min}}_{lm} =  n_d \, \sigma_0 \, \sum_{s=1}^{n_\text{D}} \frac{M^{(s)}_{lm}}{\sqrt{\det M^{(s)}}}.
\label{gen-sigma-sum:eq}
\end{equation}
This universal procedure to find the minimal conductivity for clean systems  can be applied for all Hamiltonians given
by Eq.~(\ref{gen-Ham:eq}) except for cases when the determinant of the matrix $M$ is zero at any of the Dirac points.
In this case, we cannot derive an universal and $\eta$ independent expression for the minimal conductivity and
one needs to evaluate Eq.~(\ref{sigma_Kubo:eq}) numerically for finite $\eta$.
However, if the value of $\eta$ in Eq.~(\ref{sigma_Kubo:eq}) is less than the energy scale for which the quadratic expansion of the dispersion relation is valid  
then our  expression (\ref{gen-sigma-sum:eq}) gives the same result as that obtained numerically from Eq.~(\ref{sigma_Kubo:eq}). 
In such cases the minimal conductivity is independent of the value of $\eta$, ie, the microscopic details of the systems.

\emph{Applications.}---We now apply our method for calculating the minimal conductivity in different models of bilayer graphene.

(i) The simplest case is when the trigonal warping is absent, ie, the linear term in momentum is missing in Eq.~(\ref{hp_2layer:eq}).
Then we find that $\mbox{\boldmath $\sigma$}^{min}(w)  = 8 \, \sigma_0 I_2$ and independent of $w$ and $\eta$.

\begin{figure}[hbt]
\vspace{-2mm}
\includegraphics[height=3.5cm]{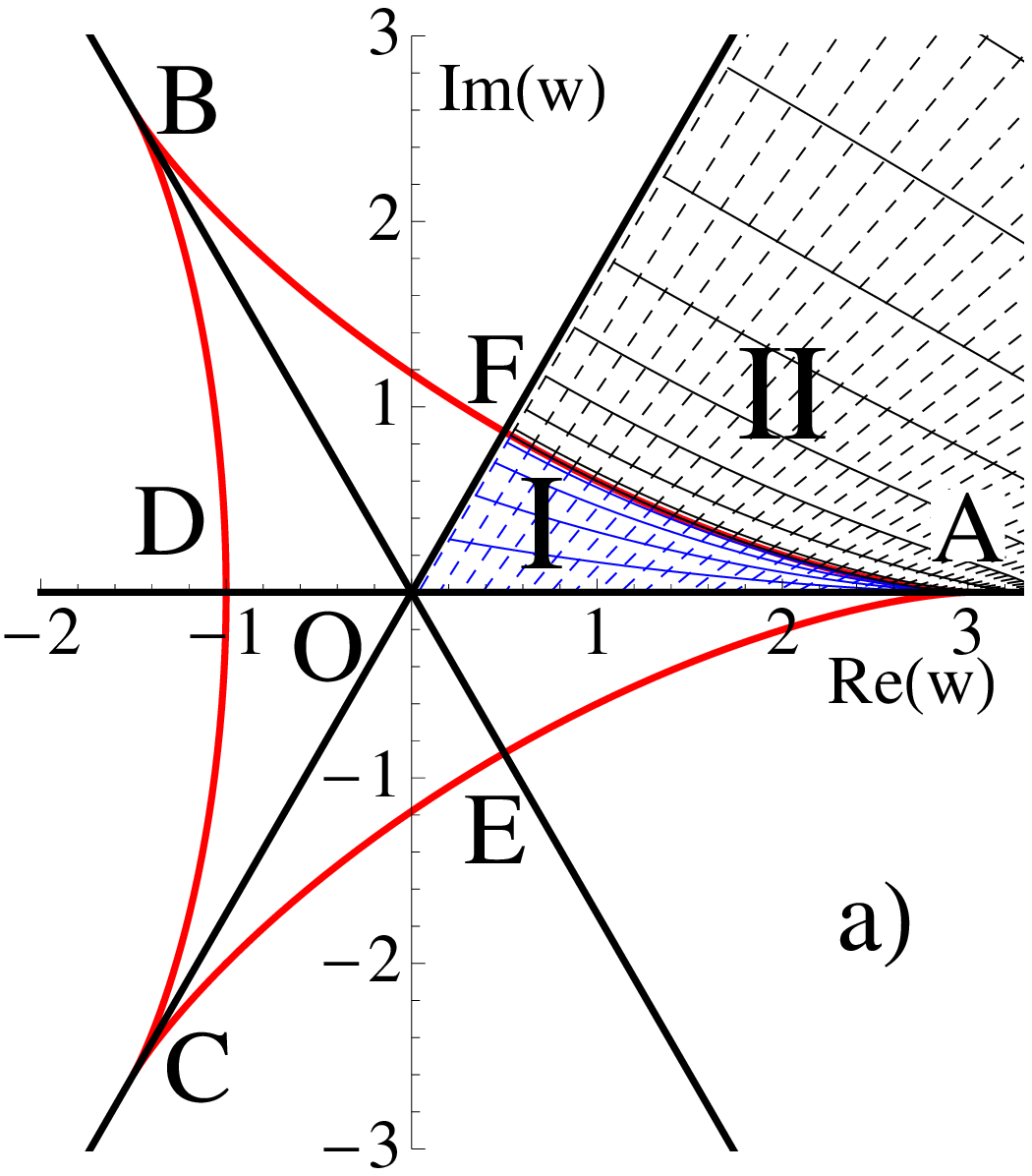} \hspace{7mm}
\includegraphics[scale=0.5]{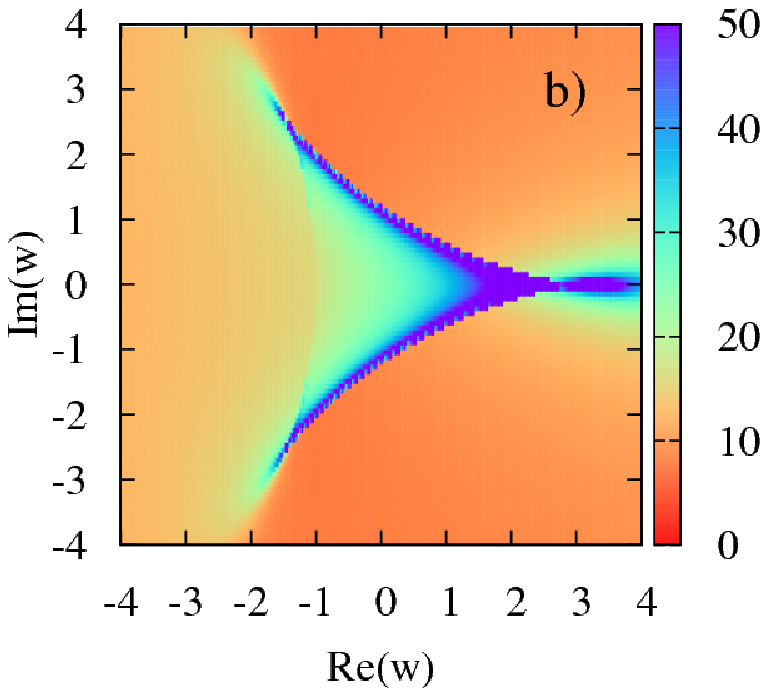} 
\includegraphics[scale=0.5]{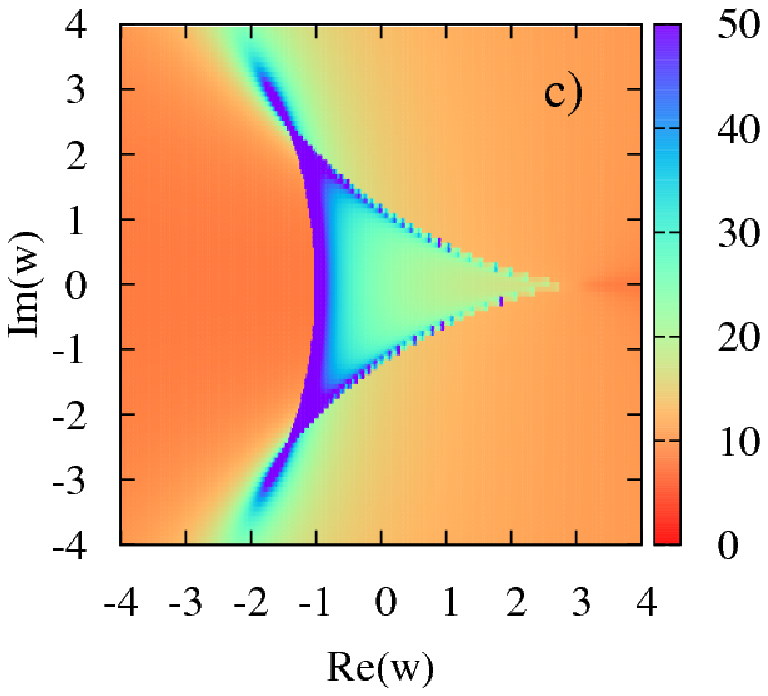}
\includegraphics[scale=0.5]{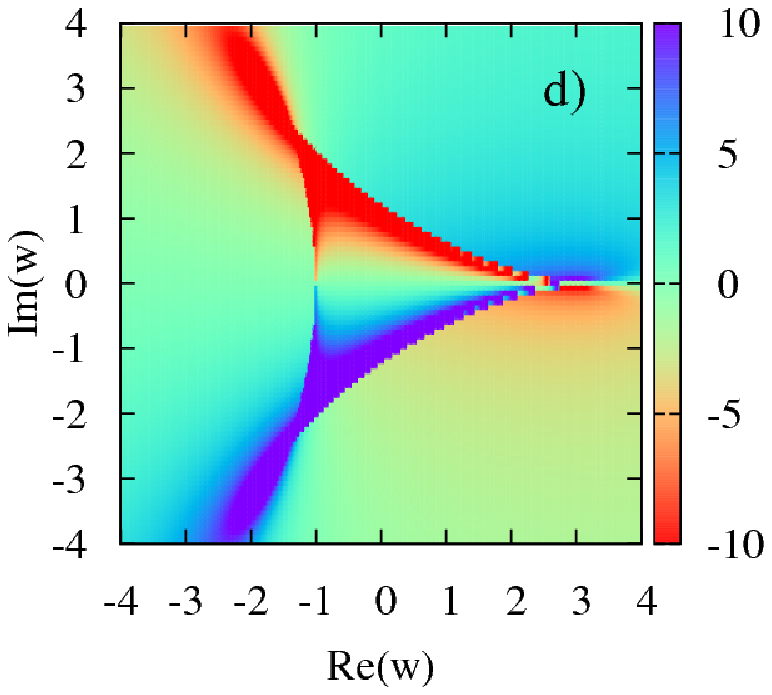}
\caption{\label{Kubo-w-plane:fig}
(Color online)  a) The complex $w$-plane with the triangle like curve ABC  $ w_{\triangle}$ (red solid line) and regions I and II.
The dashed and solid lines in regions I and II correspond to constant parameters $\alpha$ and $\beta$, respectively.
The components of the conductivity tensor (in units of $\sigma_0$) in the complex $w$-plane:
b) $\sigma_{xx}(w)$, c) $\sigma_{yy}(w)$ and d) $\sigma_{xy}(w) = \sigma_{yx}(w)$.
Here $w$ is in units of $\varepsilon_{\text{L}}$.
}
\end{figure}

(ii) Now we take into account the effect of the trigonal warping using the Hamiltonian (\ref{gen-Ham:eq})--(\ref{hp_2layer:eq}).
Then, it is easy to calculate the matrix $M$ defined in Eq.~(\ref{gen-M-matrix:eq}) and we have
\begin{eqnarray}
M (\mathbf{k}) &=& \left( \begin{array}{cc}
 1+ {\bf k}^2 -2 k_x & 2 k_y \\
2k_y & 1+ {\bf k}^2 + 2 k_x
\end{array}  \right) ,
\label{M_matrix:eq}
\end{eqnarray}
and $\det M = {( {\bf k}^2 -1)}^2$, where the wave number ${\bf k}$ is in units of $m v_3/ \hbar$ 
 and the matrix $M$ is in units of ${(\hbar \varepsilon_{\text{L}}/m v_3)}^2$.
Thus, the\emph{ singular points} (when $\det M = 0$) are located on a unit circle in the ${\bf k}$-plane. 
In this case the parameter $w$ lies on the triangular like curve ABC on the complex $w$-plane shown in Fig~\ref{Kubo-w-plane:fig}a
(the same is plotted in Fig.~3a in Ref.~\onlinecite{Falko_strain_2layer:cikk}).
At these points two Dirac cones collide and annihilate resulting in a topological phase transition in the energy band dispersion.
There are \textit{four}/\textit{two} Dirac points in the momentum plane for the parameter $w$ lying inside/outside
of the triangular like curve ABC, respectively.

In what follows it is useful to parameterize the complex parameter $w$ in region I and II shown in Fig~\ref{Kubo-w-plane:fig}a in the following way
\begin{equation}
\label{w-param:eq}
 w = e^{-i 4\alpha} + 2\cos (2 \beta) \, e^{i 2\alpha},
\end{equation}
where  $\alpha \in [0,\pi /3]$ and $0 < \beta < \text{min}\{3\alpha, \pi - 3\alpha\}] $  in region I,
while in region II $\beta$ is a pure imaginary number such that $i \beta < 0$.
The triangular like curve ABC in Fig~\ref{Kubo-w-plane:fig}a in this parametrization reads as
$ w_{\triangle}(\alpha) = e^{-i 4\alpha} + 2 \, e^{i 2\alpha}$ with $0 < \alpha \le \pi $.
The parameter $w$ lying outside regions I and II can be folded back to either of these regions by
a symmetry operation belonging to the group $C_{3v}$.
Note that if no symmetry breaking is present then $w=0$ which corresponds to $\alpha = \beta = \pi /6$.

Using the parametrization (\ref{w-param:eq}) the minimal conductivity can be analytically obtained
from Eqs.~(\ref{gen-sigma-sum:eq}) and (\ref{M_matrix:eq}) for arbitrary complex parameter $w$
(for the location of the Dirac points see the Supplementary Information).
In region I  we have
\begin{subequations}
 \label{min_cond_1-2:eq}
\begin{eqnarray}\label{mincond_inner:eq}
\lefteqn{\frac{\mbox{\boldmath $\sigma$}^{min}_I}{8\,\sigma_0} =
I_2 +\frac{1}{2\sin{\beta}\,(\sin{3\,\alpha}-\sin{\beta})}
}\nonumber \\
&\times&
\left( \begin{array}{cc}
\cos^2\alpha - \sin \beta \sin \alpha  & \cos \alpha (\sin \beta -\cos \alpha) \\
\cos \alpha (\sin \beta -\cos \alpha) & \sin \alpha (\sin \beta +\sin\alpha)
\end{array}  \right) \!\! ,
\end{eqnarray}
while in region II it is given by
\begin{eqnarray}\label{mincond_outer;eq}
\lefteqn{\frac{\mbox{\boldmath $\sigma$}^{min}_{II}}{8\,\sigma_0} =
I_2 +  \frac{1}{\cos^2{\beta}-\cos^2{3\,\alpha}}
}\nonumber \\
&\times&
\left( \begin{array}{cc}
\cos^2{2\,\alpha} & \sin{2\alpha} \cos 2\alpha \\
\sin{2\alpha} \cos 2 \alpha& \sin^2{2\,\alpha}
\end{array}  \right).
\end{eqnarray}
\end{subequations}
The eigenvalues of the conductivity tensor:
$\sigma^I_{1,2}/(8\sigma_0) = 1+2/\left(1\pm\sqrt{1-4\sin\beta\left(\sin 3\alpha - \sin\beta\right)}\right)$ in region I and
$\sigma^{II}_{1}/(8\sigma_0) =1$, while $\sigma^{II}_{2}/(8\sigma_0) =1+ 1/\left(\cos 2\beta -\cos 6 \alpha\right)$  in region II.
The minimal conductivity for $w$ lying outside regions I and II can be obtained by
$\mbox{\boldmath $\sigma$}^{min}(w) = R^{-1} \cdot \mbox{\boldmath $\sigma$}^{min}(R\cdot w) \cdot R$,
where $R$ is a symmetry operation of the symmetry group $C_{3v}$ (reflection or $120^\circ$ rotation)
which transforms $w$ into region I or II.
Therefore, the minimal conductivity has a $C_{3v}$ symmetry in the $w$-plane.
The minimal conductivity tensor ${\mbox{\boldmath $\sigma$}}^{\text{min}}$ is a symmetric matrix but generally
for complex $w$  the non-diagonal elements can be different from zero and even can be negative.
Note that for $w=0$ (without symmetry breaking) we recover our earlier results, namely
$\sigma^{\text{min}}_{xx}(w=0) = \sigma^{\text{min}}_{yy}(w=0) = 24 \, \sigma_0 $ and the off-diagonal elements are zero~\cite{cserti:066802}.

We numerically calculated the components of the minimal conductivity tensor from  (\ref{min_cond_1-2:eq})
(or from Eqs.~(\ref{gen-sigma-sum:eq}) and (\ref{M_matrix:eq}))
for complex parameter $w$ and plotted in Fig.~\ref{Kubo-w-plane:fig}b, c and d, respectively.
One can see from the figures that the points  $ w_{\triangle}$ along the curve ABC, ie, where the electronic topological transition occurs
the conductivity changes dramatically.
We compared our theoretical prediction (\ref{min_cond_1-2:eq}) with that obtained numerically from~(\ref{sigma_Kubo:eq})
for finite $\eta$ and found that the numerical results becomes better with decreasing $\eta$
(for details see Fig.~\ref{min_s_xx_yy_u_num:fig} in the Supplementary Information).
The analytical result~(\ref{min_cond_1-2:eq}) starts to deviate from the numerical calculations at some values of parameter $w$
for which the Dirac points are close enough that the energy corresponding to the saddle point between them becomes
the same order of magnitude as the inverse lifetime $\eta$.

\begin{figure}[hbt]
\includegraphics[scale=0.63]{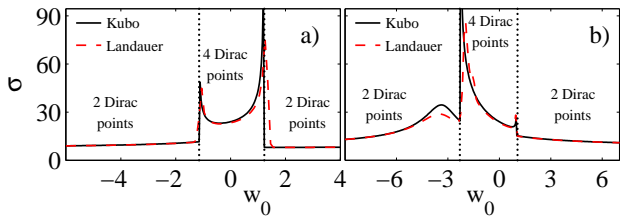}
\caption{\label{Kubo-Landauer:fig}
(Color online) The conductivity $\sigma_{\text{Kubo}}(w_0)$ (black solid line) and by numerical evaluation of the Landauer formula (red dashed line)
as functions of the parameter $w_0$, where  $w=w_0 e^{i\omega}$, and a)  $\omega =27^\circ$, $\theta = -42^\circ$
and b) $\omega =57^\circ$, $\theta = 42^\circ$.
The units are the same as in Fig.~\ref{Kubo-w-plane:fig}.
The dotted vertical lines indicate the values of the parameter $w_0$ where the number of Dirac points changes.
}
\end{figure}

To confirm our analytical predictions (\ref{min_cond_1-2:eq}) we calculated the conductance using the two-terminal Landauer formula 
(for details see the Supplementary Information).
In these calculations the orientation ${\bf n}=(\cos \theta ,\sin \theta ) $ of the graphene sample with respect to the contacts is fixed 
(see Fig.~\ref{fig_geometry} in the Supplementary Information).
From the conductivity tensor~(\ref{min_cond_1-2:eq}) the conductivity along the direction ${\bf n}$ in the two-terminal measurement is given by
$\sigma_{\text{Kubo}}(w) = {\bf n}\cdot  \mbox{\boldmath $\sigma$}(w) \cdot {\bf n}$
~\cite{PhysRevB.40.8169,Ryu-Ludwig:cikk_PhysRevB.75.205344}.
Then this results should be compared with that obtained numerically from the Landauer formula.
Our results are shown in Fig.~\ref{Kubo-Landauer:fig}.
The complex parameter $w$ is taken along two different lines given by the direction $\omega$
in Fig~\ref{Kubo-w-plane:fig}a such that  $w=w_0 e^{i\omega}$, where $w_0$ varies.
Similarly, two electrode directions ${\bf n}$ are taken.
As can be seen from the Fig.~\ref{Kubo-Landauer:fig}) the agreement between the two approach is very good
for most of the values of $w_0$ (and similarly good agreements were found for other values of $w$ and directions ${\bf n}$
not presented here).
Note that the electronic topological transition occurs at two values of $w_0$, where the number of Dirac points
changes from four to two (see the dotted lines in Fig.~\ref{Kubo-Landauer:fig}).
Near the singular points the deviation between the numerical and analytical results arises from the fact
that here the Dirac cones come closer to each other than the momentum space resolution of the numerical method
dictated by the finite size of the sample in the Landauer approach.
Thus, not too close to the singular points $w_{\triangle}$ our central result~(\ref{min_cond_1-2:eq}) well approximates 
the elements of the conductivity tensor for bulk and clean bilayer graphene which is independent of the inverse lifetime $\eta$ and  
depends on the topology of the Fermi surface at the neutrality point.

We now comment the experimental feasibility of measuring the minimal conductivity.
On the one hand, in recent experiments~\cite{Novoselov_private_comm} the bilayer graphene samples are clean enough
that it is ballistic up to lengths $2-4\;\rm{\mu m}$ making sense to apply the Landauer approaches and to compare with our analytical result~(\ref{min_cond_1-2:eq}) obtained from the Kubo formula.
Furthermore, the temperature experimentally can be as low as $T \approx 0.25 \, \text{K} \approx 0.02 $ meV, ie,
its effect can be neglected in first approximation.
On the other hand, the conductivity depends very sensitively on  the orientation of the sample
and the values of the parameter $w$.
Thus, these two unknown factors seem to be crucial to reproduce the measured minimal conductivity.
However, with the experimental control of the direction of the electrodes the measurement of the minimal conductivity
may provide a good tool for determining the complex parameter $w$ and exploring its origin
in the electronic topological transition.

In conclusion, using the Kubo formula we calculated analytically the minimal conductivity in bilayer graphene
taking into account the electronic topological transition.
Our results are confirmed by numerical calculations of the conductivity using the Landauer approach.
We hope that our analysis may provide a better insight into the origin of the reconstruction of the electronic spectrum
observed in recent experiment~\cite{Novoselov_private_comm}.
Note that our general approach for calculating the minimal conductivity can be applied to various other electronic systems.

We gratefully acknowledge the fruitful discussions with K. S. Novoselov, V. P. Gusynin and A. P\'alyi.
This work was partially supported by the Hungarian Science Foundation OTKA under the contracts No. 75529 and No. 81492,
by the Marie Curie ITN project NanoCTM (FP7-PEOPLE-ITN-2008-234970),
by the European Union and co-financed by the European Social Fund (grant agreement no. TAMOP 4.2.1/B-09/1/KMR-2010-0003).


\clearpage

\begin{center}
{\bf Supplementary Information: }

\bigskip

{\bf Effect of the band structure topology on the minimal conductivity for bilayer graphene with symmetry breaking }
\vspace{5mm}

Gyula D\'avid$^1$, P\'eter Rakyta$^{2}$, L\'aszl\'o Oroszl\'any$^2$ and  J\'ozsef Cserti$^2$

\vspace{5mm}
$^1$\textit{ Department of Atomic Physics, E{\"o}tv{\"o}s University, H-1117 Budapest, P\'azm\'any P{\'e}ter s{\'e}t\'any 1/A, Hungary. }\\
$^2$\textit{Department of Physics of Complex Systems, E{\"o}tv{\"o}s University,  H-1117 Budapest, P\'azm\'any P{\'e}ter s{\'e}t\'any 1/A, Hungary}.


\bigskip

\end{center}

\supplementPRL

\section{Dirac points}

In the parametrization of the parameter $w$ given by Eq.~(\ref{w-param:eq}) in the main text 
the zeros ${\bf k}^{(s)}$ in region I are given by
\begin{subequations}
\label{4-2_Dirac_points:eq}
\begin{eqnarray}
\label{kI1:eq}
{\bf k}^{(1)} &=& \left( \begin{array}{c}
 \cos{2\,\alpha}+2\,\cos{\beta}\,\cos{\alpha} \\[1ex]
 \sin{2\,\alpha}-2\,\cos{\beta}\,\sin{\alpha}
\end{array}  \right),
\end{eqnarray}
\begin{eqnarray}
\label{kI2:eq}
{\bf k}^{(2)} &=& \left( \begin{array}{c}
 \cos{2\,\alpha}-2\,\cos{\beta}\,\cos{\alpha}   \\[1ex]
 \sin{2\,\alpha}+2\,\cos{\beta}\,\sin{\alpha}
\end{array}  \right),
\end{eqnarray}
\begin{eqnarray}
{\bf k}^{(3)} &=& \left( \begin{array}{c}
  -\cos{2\,\alpha}+2\,\sin{\beta}\,\sin{\alpha} \\[1ex]
 -\sin{2\,\alpha}+2\,\sin{\beta}\,\cos{\alpha}
\end{array}  \right),
\end{eqnarray}
\begin{eqnarray}
{\bf k}^{(4)} &=& \left( \begin{array}{c}
  -\cos{2\,\alpha}-2\,\sin{\beta}\,\sin{\alpha}  \\[1ex]
  -\sin{2\,\alpha}-2\,\sin{\beta}\,\cos{\alpha}
\end{array}  \right),
\end{eqnarray}
\end{subequations}
while in region II there are only two Dirac points and they are the same as Eqs.~(\ref{kI1:eq}) and (\ref{kI2:eq}) but with purely imaginary $\beta$.

\section{Real parameter $w$}

For real parameter $w$, ie, for  $w =u$, where $u \in \mathbb{R}$ we can give analytical expressions 
for the components of the minimal conductivity tensor directly in terms of the real parameter $u$ 
(without using the parameters $\alpha$ and $\beta$).
In this case, the zeros ${\bf k}^{(s)}$ as functions of $u$ are given by
\begin{subequations}
\begin{eqnarray}
\label{ks_gyok:eq}
{\bf k}^{(1)} &=&  {(-1,-\sqrt{3-u})}^{T}, \quad \text{if \, $u \le 3$},   \\[1ex]
{\bf k}^{(2)} &=&  {(-1,\sqrt{3-u})}^{T}, \quad \text{if \, $u \le 3$},   \\[1ex]
{\bf k}^{(3)} &=&  {(1-\sqrt{u+1},0)}^{T}, \quad \text{if \, $u \ge -1$},   \\[1ex]
{\bf k}^{(4)} &=&  {(1+\sqrt{u+1},0)}^{T}, \quad \text{if \, $u \ge -1$},
\end{eqnarray}
\end{subequations}
where $T$ denotes the transpose operation.
For $u=0$ no symmetry breaking occurs in bilayer graphene, and the zeros  at ${\bf k}^{(2)} = (0,0)$
and the remaining three ones are located at the corner of a regular triangle as well known.
For the singular point $u=-1$ the zeros ${\bf k}^{(3)}$ and ${\bf k}^{(4)}$ collide at ${\bf k}=(1,0)$, while
for $u=3$ the zeros ${\bf k}^{(1)}, {\bf k}^{(2)} $ and ${\bf k}^{(3)}$ collide at ${\bf k}=(-1,0)$.

Evaluating the matrix $M$ at the zeros ${\bf k}^{(s)}$ ($s=1,\dots,n_{\text{D}}$) and using Eq.~(\ref{gen-sigma-sum:eq}) in the main text 
we find (after a simple algebra) that the universal minimal conductivity in clean bilayer graphene
as a function of the parameter $u$ is given by
\begin{subequations}
\label{s_xx_yy_u:eq}
\begin{eqnarray}
\sigma^{\text{min}}_{xx}(u) &=& \sigma_0 \, \begin{cases}
 \frac{8\left(7-u \right)}{3-u}, & \, \text{if \, $u<-1$},   \\[1ex]
\frac{8\left(7-u+2\sqrt{u+1}\right)}{3-u}, & \,   \text{if \, $-1 \leq u < 3$}, \\[1ex]
\frac{8\left(u+1\right)}{u-3}, & \,  \text{if \, $u > 3$}.
\end{cases}
\end{eqnarray}
\begin{eqnarray}
\label{s_yy_u:eq}
\sigma^{\text{min}}_{yy}(u) &=& \sigma_0 \, \begin{cases}
8, & \, \text{if \, $u<-1$},   \\[1ex]
8 + \frac{16}{\sqrt{u+1}}, & \, \text{if \, $-1 < u < 3$}, \\[1ex]
8, & \,  \text{if \,  $u > 3$},
\end{cases}
\end{eqnarray}
\end{subequations}
and $\sigma_{xy}(u) = \sigma_{xy}(u) = 0$ for all values of $u$, ie,
the conductivity tensor $\mbox{\boldmath $\sigma$}$ is a diagonal matrix.
Note that for $u < -1$  or $ u >3$, ie, where only two Dirac points exist $\sigma^{\text{min}}_{yy}(u)$ is constant.

\section{Finite inverse lifetime}

\begin{figure}[hbt]
\includegraphics[scale=0.63]{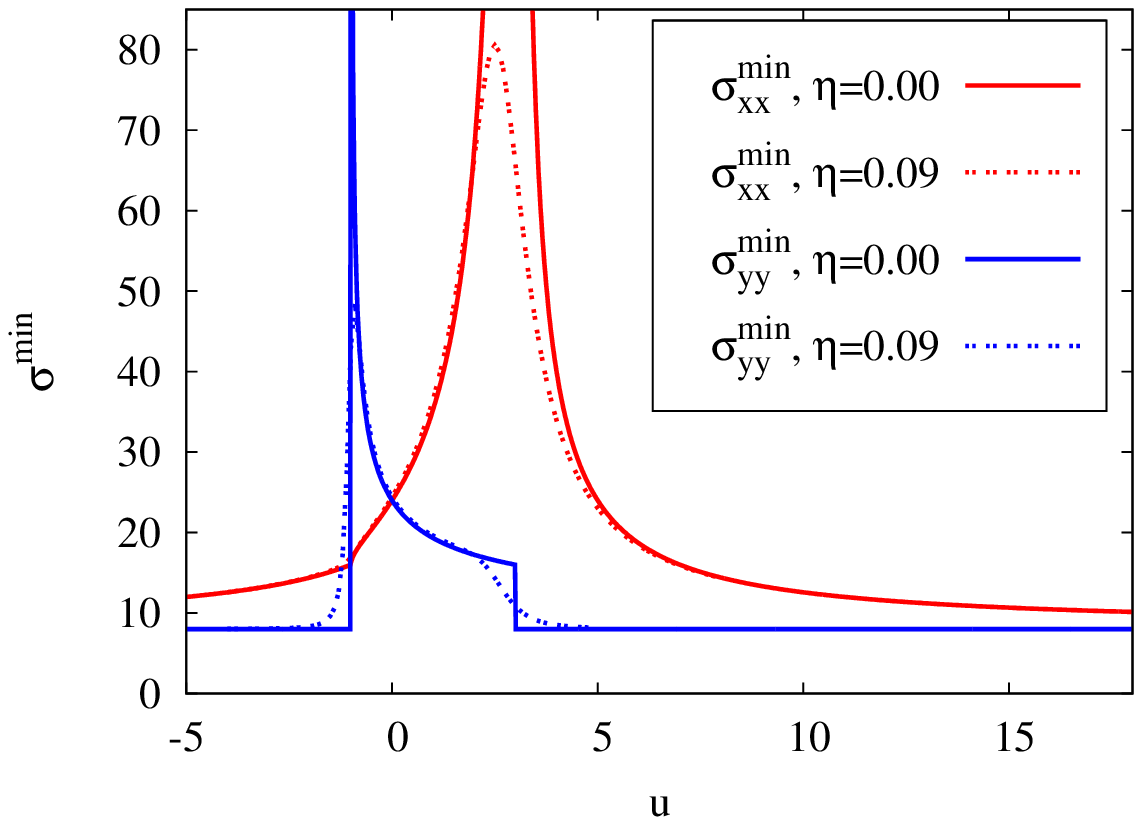}
\caption{\label{min_s_xx_yy_u_num:fig}
(Color online) The diagonal components of the minimal conductivity tensor as functions of the real parameter $u$
obtained from (\ref{s_xx_yy_u:eq}) (solid lines)
and from numerical evaluation of Eq.~(\ref{sigma_Kubo:eq}) in the main text (dashed lines) corresponding to  $\eta = 0.09 $.
The units are the same as in Fig.~\ref{Kubo-w-plane:fig} in the main text.
For $-1< u <3$ there are four Dirac cones, while for $u<-1$ or $u>3$ only two Dirac cones exist.
}
\end{figure}

For real parameter $u$ our analytical result (\ref{s_xx_yy_u:eq}) is compered with the numerical results obtained
from (\ref{sigma_Kubo:eq}) in the main text and plotted in Fig.~\ref{min_s_xx_yy_u_num:fig}.
We found that the agreement between the theoretical predictions~(\ref{s_xx_yy_u:eq}) and the numerical results
becomes better with decreasing $\eta$.
The reason for the deviation between the analytical~(\ref{s_xx_yy_u:eq}) and numerical results is discussed in the main text.

\section{Landauer's approach for calculating the conductivity }

\begin{figure}
 \centering
 \includegraphics[width=0.4\textwidth]{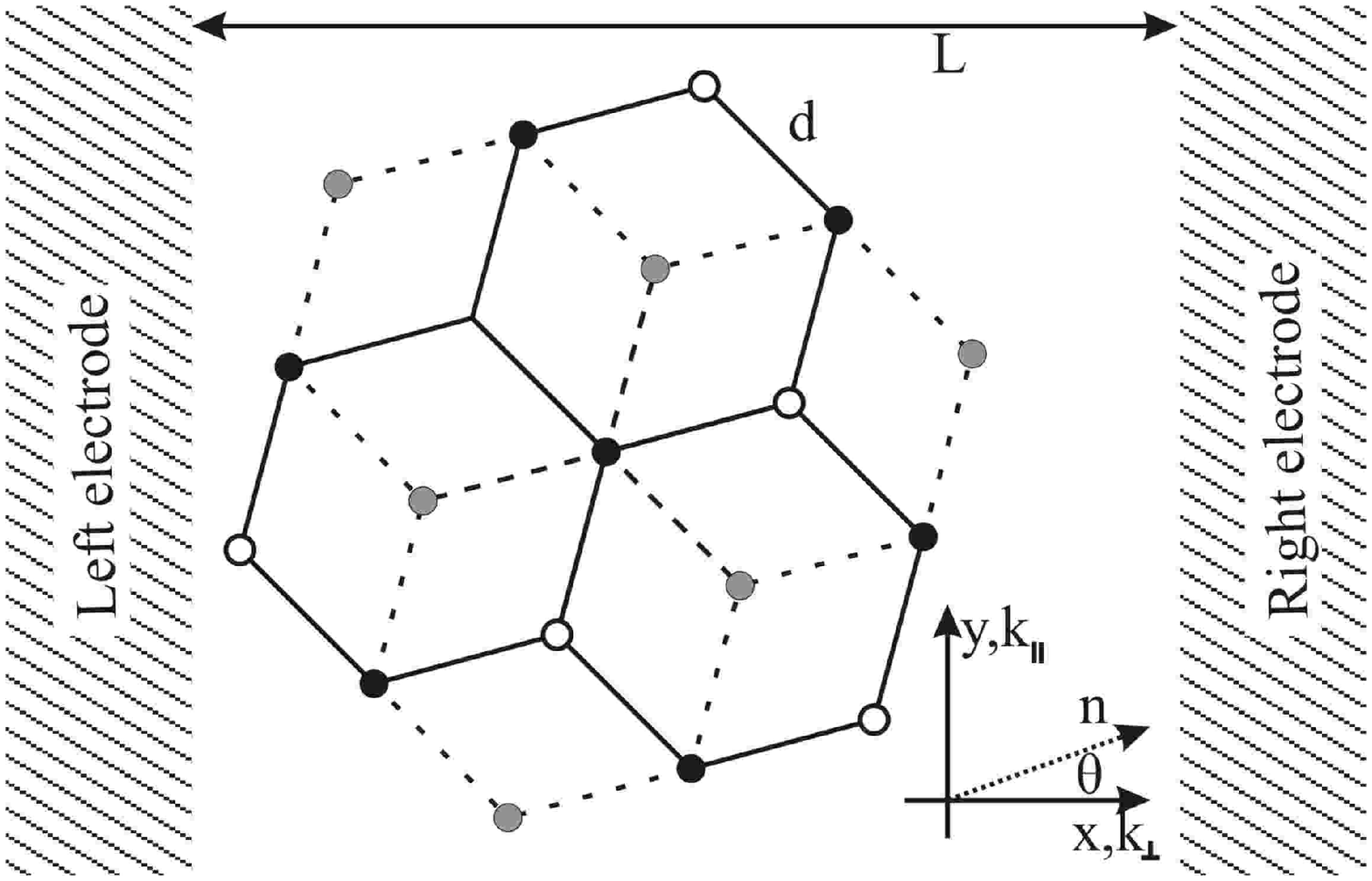}
 \caption{Schematic of a wide bilayer strip of length $L$, contacted by two electrodes.
Solid lines in the lattice represent the upper layer, while the dashed lines the lower layer.
The electrodes are at an angle $\theta$ with respect to the symmetry axis of the bilayer lattice denoted by ${\bf n}$.
The wave vector $\mathbf{k_{\perp}}$ ($\mathbf{k_{\parallel}}$) is perpendicular (parallel) to the electrodes.
The carrier concentration around the neutrality point can be tuned by a separate gate electrode not shown in the figure.}
\label{fig_geometry}
\end{figure}

Landauer's approach, in a similar way as in Refs.~\cite{Katsnelson_Klein:ref,PhysRevLett.96.246802,PhysRevB.79.073401_S},
gives the conductance in terms of transmission probabilities for an electron to be transmitted by a sample.
In our two-terminal calculations, we employ a model of infinite wide bilayer strip of length $L$ with carrier concentration near the charge neutrality point as
the scattering region, and two highly doped regions as the electrodes
oriented at angle $\theta$ with respect to the symmetry axis of the bilayer lattice (see Fig.~\ref{fig_geometry}).
According to recent experiments~\cite{Novoselov_private_comm_S} in very clean samples we can consider the bilayer strip
to be ballistic up to lengths $2-4\;\rm{\mu m}$.
In our numerical calculations we chose $L=2.5\;\rm{\mu m}$.
The connected electrodes are modeled by semi-infinite bilayer sheets under negative bias voltage $V_0$  much higher than any other characteristic energy scale
in the system.
Since properties of the transmission of the strip saturate for $V_0<-1\;\rm{eV}$, in our numerical calculations we optionally applied $V_0=-10\;\rm{eV}$.
To describe low energy excitations around the charge neutrality point in the vicinity of the Brillouin zone corners
$\mathbf{K}$ ($\xi=1$) and $\mathbf{K'}$ ($\xi=-1$), according to Refs.~\cite{Falko_strain_2layer:cikk_S,PhysRevB.82.201408_S},
we consider the Hamiltonian:
\begin{equation}
 H_4(\mathbf{k}) = \begin{pmatrix}
      0 & \xi v_3\hbar k_+ + w & 0 & \xi v\hbar k_- \\
      \xi v_3\hbar k_- + w^* & 0 & \xi v\hbar k_+ & 0 \\
      0 & \xi v\hbar k_- & 0 & \gamma_1 \\
      \xi v\hbar k_+ & 0 & \gamma_1 & 0
     \end{pmatrix}\;, \label{eq:H44}
\end{equation}
where $k_{\pm} = e^{\mp i\theta}(k_{\perp}\pm \rm{i}k_{\parallel})$
(here $k_{\perp}$/$k_{\parallel}$ is the component of the wave vector perpendicular/parallel to the electrodes),
$v=3/2 \, d \gamma_0/\hbar$ is the Dirac velocity in the monolayer,
$v_3=3/2 \,d \gamma_3/\hbar$, and  $d$ is the bond length between the carbon atoms,
and finally $w$ is the same parameter than in the $2\times 2$ Hamiltonian given
by Eqs.~(\ref{gen-Ham:eq}) and (\ref{hp_2layer:eq}) in the main text.
Here we take $\gamma_0\sim 3\;\rm{eV}$, $\gamma_1\sim 0.4\;\rm{eV}$ and $\gamma_3\sim 0.3\;\rm{eV}$ for the intra- and interlayer
hopping parameters, respectively~\cite{Falko_strain_2layer:cikk_S}.
In bulk systems the low energy band structure is well approximated by a two by two effective Hamiltonian
when the high energy dimer states are decimated~\cite{mccann:086805_S}.
However, in this case this decimation cannot be used directly since a potential step generates further terms in the effective Hamiltonian.

The band structure around $\mathbf{K}$ and $\mathbf{K'}$ are identical up to a rotation by $\pi$ in momentum space.
Since the conductance is invariant under this transformation, it is sufficient to perform the calculations only at $\mathbf{K}$ point,
since the other valley would lead to the same contribution
to the conductance.

The transmission probabilities is obtained by solving the scattering problem.
Electronic states are specified by their energy $\varepsilon$ and the transverse momentum $k_{\parallel}$, which are conserved in the scattering process.
For a given $\varepsilon$ and $k_{\parallel}$ there are four solutions of longitudinal wave vector $k_{\perp}$ which satisfies the characteristic equation
$\det\left[H_4(k_{\perp}, k_{\parallel}) - I_4 \, \varepsilon\right] = 0$, where $I_4$ is the $4\times 4$ identity matrix.
Solutions $k_{\perp}^{(n)}$ ($n\in\{1,\dots ,4\}$) can be obtained if one rearranges the characteristic polynomial into the form:
\begin{equation}
 \det\left[H_4(k_{\perp}, k_{\parallel}) - I_4 \, \varepsilon\right] = \sum\limits_{m=0}^4 a_m(\varepsilon, k_{\parallel})k_{\perp}^m\;.
\label{eq:polynom}
\end{equation}
The complex roots $k_{\perp}^{(n)}$ of the polynomial (\ref{eq:polynom}) are equal within numerical precision to the eigenvalues of the companion matrix of
this polynomial~\cite{companion:book}.
Electronic states in the scattering region ($0\leq x\leq L$) are denoted
by $\Psi_{\text{sc}}^{(n)}(x,y) = \Phi_{\text{sc}}^{(n)} e^{i(k_{\perp}^{(n)}x+k_{\parallel}y)}$,
where $\Phi_{\text{sc}}^{(n)}$ satisfy relation for all possible $n$:
\begin{equation}
 H_4(k_{\perp}^{(n)}, k_{\parallel})\, \Phi_{\text{sc}}^{(n)} = \varepsilon \, \Phi_{\text{sc}}^{(n)}.
\label{eq:eigvec}
\end{equation}
The scattering state in the bilayer strip is then a linear combination of these four electronic states.
Similarly, in the electrodes ($x<0$ or $x>L$), one can obtain the longitudinal wave numbers $k_{\perp, \text{lead}}^{(n)}$ and the corresponding electronic states
$\Psi_{\text{lead}}^{(n)}(x,y) = \Phi_{\text{lead}}^{(n)}\, e^{i(k_{\perp,\text{lead}}^{(n)}x+k_{\parallel}y )}$ using Eqs.~(\ref{eq:polynom}) and (\ref{eq:eigvec})
with substitution $\varepsilon\rightarrow\varepsilon-V_0$.
Generally,  inside each electrode two states are propagating or decaying to the left (labeled by $L$) and other two to the right (labeled by $R$).
If we assume two incident states in the left electrode, than the $i$th ($i=1,2$) scattering state takes form:
\begin{equation}
 \Psi^{(i)}(x,y) = \left\{\begin{array}{ll}
 \Psi_{\text{lead}}^{(R_i)}(x,y) + \sum\limits_{n=1}^2r_n^i\Psi_{\text{lead}}^{(L_n)}(x,y), & x<0 , \\
 \sum\limits_{n=1}^4A_n\Psi_{\text{sc}}^{(n)}(x,y), & 0\leq x\leq L ,\\
 \sum\limits_{n=1}^2t_n^i\Psi_{\text{lead}}^{(R_n)}(x,y), & L<x ,
 \end{array}
  \right.
\end{equation}
where $r_n^i$ and $t_n^i$ are the reflection and transmission amplitudes, which have to be determined (together with coefficients
$A_n$) by imposing the continuity condition of the wave functions at the boundaries $x=0$ and $x=L$.
Then the conductivity, including valley and spin degeneracies, reads~\cite{PhysRevB.79.073401_S}:
\begin{equation}
 \frac{\sigma_{\text{Landauer}}}{\sigma_0} = 2 L\int\limits_{-\infty}^{\infty}{\rm d}k_{\parallel}\sum\limits_{n=1}^2\sum\limits_{i=1}^2 \left|t_n^i\right|^2,
\label{eq:G}
\end{equation}
where $\sigma_0 = e^2/(\pi h)$.
In the numerical calculations we find that the conductivity $\sigma$ is practically independent of $L$
in the ballistic regime with lengths $L = 2-4\;\rm{\mu m}$.

Finally, we note that
 (i) the conductivity calculated from Eq.~(\ref{eq:G}) becomes sensitive to peculiar dynamics of the nematic phase transition when
 the length of the bilayer strip ($L$) is much larger than the characteristic length of the band structure
$l\sim \pi/\Delta k$, where $\Delta k$ is the splitting of the Dirac cones.
For $w=0$ the characteristic length of the band structure is $l\sim 50\;{\rm nm}$~\cite{PhysRevB.79.073401_S}, which is much smaller than the
 length of the bilayer strip $L=2.5\;{\rm \mu m}$ used in our calculations,
 (ii) when two Dirac cones align  in the direction $\mathbf{k}_{\perp}$, the resulting conductivity will show
quantum interference of electronic states of the two cones.

\end{document}